\documentstyle[preprint,aps]{revtex}
\tighten
\begin{document}
\draft
\preprint{UCF-CM-93-005}
\title{Haldane fractional statistics in the fractional quantum Hall effect}
\author{M.D. Johnson$^{*}$ and G.S. Canright$^{\dag}$}
\address{
$^{*}$Department of Physics, University of Central Florida, Orlando, FL
32816-2385\\
$^{\dag}$Department of Physics. University of Tennessee, Knoxville, TN
37996-1200}
\maketitle
\begin{abstract}
We have tested Haldane's ``fractional-Pauli-principle'' description of
excitations around the $\nu = 1/3$ state in the FQHE,
using exact results for small systems of
electrons.
We find that Haldane's prediction
$\beta = \pm 1/m$ for quasiholes and quasiparticles, respectively,
describes our results well with the modification
$\beta_{qp} = 2-1/3$ rather than
$-1/3$.
We also find that this approach enables us to better understand the {\it
energetics\/} of the ``daughter'' states; in particular, we find good
evidence, in terms of the effective interaction
between quasiparticles, that the states $\nu = 4/11$ and 4/13 should not be
stable.
\end{abstract}
\pacs{05.30.-d,73.20.Dx,73.20.Mf}

\narrowtext

The traditional way to define fractional statistics is in terms of
the exchange phase $e^{i\alpha\pi}$ which is developed by the wavefunction
when two identical particles are exchanged \cite{AnyonReview}.
For bosons $\alpha=0$
and for fermions $\alpha=1$.  Particles with fractional statistics
({\it i.e.}, fractional values of $\alpha$) can, in principle, exist
in two dimensions.  A common model of fractional statistics
represents this exchange phase by adding an infinitesimal tube of
magnetic flux (and a coupling charge) to ordinary fermions or bosons.

Recently Haldane \cite{HaldaneFS}
proposed an alternative approach to fractional statistics which applies
to the case of systems with a finite Hilbert space (or subspace).
Letting $d_n$ be the size of the single-particle Hilbert space available
to the $n^{th}$ identical particle, Haldane defined a fractional
exclusion coefficient $\beta$ by $\beta = -(d_{n+\Delta n}-d_n)/\Delta n$.
Haldane argued for a correspondence between the exchange phase $\alpha$
and the generalized Pauli coefficient $\beta$ for the bulk excitations
in the fractional quantum Hall effect (FQHE) \cite{MacDonaldBook},
based on the variational form of their wavefunctions.
For quasiholes (qh) and quasiparticles (qp) near filling factor $\nu=1/m$,
with $m$ an odd integer, Haldane obtained $\beta_{qh}=1/m$
and $\beta_{qp} = -1/m$.

In this paper we test the utility of Haldane's approach for describing
the low-energy excitations in the FQHE, using exact diagonalization of
small systems of electrons. Our method
relies on the single assumption that the number of
quasiholes and quasiparticles in a many-electron state
may be counted in the usual way.
We first obtain the many-particle generalization of Haldane's
definition, which is needed for analyzing the numerical results.
We then find that Haldane's description works well---giving an
unambiguous determination of the fractional exclusion coefficient $\beta$
for a given energy scale---with one
significant modification, namely, that $\beta_{qp}$ is positive.
Our analysis yields as a further result
information about the effective interactions between the excitations,
in terms of which the absence of the fractions $\nu=4/13$ and $\nu=4/11$ can
be simply understood.

The belief that bulk excitations in the FQHE should exhibit fractional
($\alpha$) statistics \cite{Arovas} is based upon
the variational forms for their wavefunctions.  These variational
forms accurately describe the numerically-obtained eigenstates
for a single qh or qp \cite{LaughlinBook} or for
several \cite{YangSu}.
The variational wavefunctions develop phases under the motion of a
qh or qp exactly as if it saw the electrons as quanta of flux \cite{Arovas}.
This is the basis of the `hierarchy' explanation of the allowed fractions
in the FQHE \cite{HalperinHierarchy,HaldaneHierarchy}.  An alternative
construction of the bulk excitations is given by a mapping to the integer
quantum Hall effect \cite{MacDonaldMap}.
This has recently been used by Jain and co-workers to construct
other variational wavefunctions as
an alternative explanation of the FQHE \cite{Jain}.

In this paper we are interested in analyzing the structure of the
exact low-energy spectrum (obtained numerically).
In the case of quasiholes it was first shown by Johnson and
MacDonald \cite{JohnsonMacDonald} that the nominal qh states are
separated by an energy gap from other states, and that these agree
in total number and in multiplet structure with either the `hierarchy'
or `integer-mapping' approach.   More recently He, Xie and Zhang
pointed out that there is a similar gap, although less pronounced,
for nominal qp states \cite{He};  in this case the low-energy set of qp states
can be thought of as those with an extra two zeros in their relative
pseudo-wavefunction.  Here too the multiplet structure and energies can
be obtained either by the hierarchy \cite{YangSu} or
integer-mapping \cite{Jain} approaches.  In this paper we show
that the exact low-energy spectra of interacting electrons
can be described simply and usefully by appealing to Haldane's formulation
of fractional statistics.

We begin by proposing a many-particle generalization of Haldane's
definition \cite{JohnsonMacDonald}. Let $N_n$ be the dimension of the
many-particle Hilbert space for $n$ identical particles.
($N_n$ is taken finite, by assuming a finite volume and an energy cutoff
\cite {HaldaneFS}.)
Then
\begin{equation}
N_n = { d_n + n-1 \choose n} = {d_1 + (1-\beta)(n-1) \choose n} \ \ .
\label{Nn}
\end{equation}
One can argue that Eq.~(\ref{Nn}) is no more than a restatement
of Haldane's definition of the coefficient $\beta$, via the
single-particle dimension $d_n$. An alternative derivation,
which is both illustrative and appropriate to the problem at hand,
may be made
by considering the Hilbert space of a collection of $\alpha$-anyons,
in the lowest Landau level \cite{JohnsonCanright}, confined to
a disc of radius $R$.
Wavefunctions for this problem have the form
\begin{equation}
\Psi' = f(z_1,z_2,\dots,z_n)
      \prod_{i<j=1}^n (z_i-z_j)^{\alpha} e^{-\sum_k^n |z_k|^2/4} ,
\label{wf}
\end{equation}
where $f$ is any symmetric polynomial.
One then finds, by considering the degree of $f$ in terms of the
analogous problem for bosons, that the number of states in the disc
(and in the lowest Landau level) is
${R+1 + (1-\alpha)(n-1)\choose n}$.
This is of the form Eq.~(\ref{Nn}), with $\beta$ equal to $\alpha$.
The key ingredient of this result is that increased angular momentum,
in a magnetic field and in the lowest Landau level, requires more area---and
hence fills the Hilbert space faster for fixed area.
Thus, for particles in the lowest Landau level, there is an intimate connection
between the Haldane coefficient $\beta$ and the exchange phase $\alpha$
(as noted by Haldane \cite{HaldaneFS}), namely, $\alpha = \beta
\pmod{2}$.

Consider in this example the smaller subset obtained by keeping only
those states with two extra relative zeros, {\it i.e.}, for which the
symmetric polynomial can be written with an extra Jastrow factor,
$f=\prod_{i<j}(z_i-z_j)^2f'$.
By the above argument, this extra angular momentum requires
increased area and hence fills the chosen subspace more rapidly.
In general, the subset with $2l$ extra zeros in the relative wavefunction
can be described by a Haldane coefficient
$\beta+2l$, and has a dimension
\begin{equation}
N_n^{2l} = {d_1 + (1-\beta-2l)(n-1) \choose n}.
\label{Nnl}
\end{equation}
This correspondence between extra
zeros and the value of the Haldane $\beta$ coefficient will prove to be
very useful in analyzing FQHE results.

Now we use these ideas to investigate the utility of
Haldane's approach to fractional statistics.
To test for $\beta$ statistics
we work in a
spherical geometry \cite{Sphere}, where $N_e$ electrons move in a radially
outward uniform magnetic field with $N_L-1$ flux quanta, giving
$N_L$ states in the lowest Landau level.  When $N_L=m(N_e-1)+1$, the
filling factor in the ground state is exactly $\nu=1/m$.  When $n$
extra flux quanta are added, we assume that $n$ qh are created;
or when $n$ are removed, $n$ qp are present.  One can instead create
qh/qp by keeping the Hamiltonian fixed (holding $N_L$ fixed)---which
is necessary in the determination of $\beta$---and instead
changing $N_e$.  This creates qh or qp in multiples of $m$.
We use the standard decomposition \cite{Pseudo}
for the interaction between electrons in the lowest Landau level,
resolving it into pseudopotential parameters $V_1, V_3, V_5,\dots$,
where $V_l$ gives the energy of a pair of
electrons in relative angular momentum (RAM) $l$.  These fall off
monotonically (if slowly) as $l$ grows.

In this work we will mostly consider states near $\nu=1/3$, with $V_1$
and $V_3$ nonzero, and $V_3<<V_1$.
The fractional quantum Hall effect at $\nu=1/3$ is related to the
existence of states for $\nu\leq 1/3$ which
completely avoid unit RAM, and hence pay no $V_1$ cost.
These states do, however, typically have electron pairs in RAM 3, and
hence have energies on a scale set by $V_3$.  Thus the spectrum (for
$\nu<1/3$) has a subset of low-energy (qh) states, lying below a gap
of order $V_1$
and separated among themselves by energies of order $V_3$.
The qp side ($\nu\geq1/3$) is similar but energetically harder to
separate.  We will begin by discussing the qh case.

We find numerically that when there are nominally $n$ qh present
({\it i.e.}, when $N_L=3N_e-2+n$), there is a well-defined subset of
states of dimension
\begin{equation}
N_n^{qh} = { N_e + n \choose n}
\label{Nnqh}
\end{equation}
lying below a gap in energy of order $V_1$.  This is the case for
$N_e$ from 2 to 7 and $n$ from 1 to 6.  It is illustrated for 6 electrons
in Fig.~\ref{spectrum}(a).
This dimension, and the multiplet structure, can be
explained \cite{JohnsonMacDonald} equally
well either by mapping to bosons as in the bosonic hierarchy
approach \cite{HaldaneHierarchy}, or by mapping to fermions
as in the integer-mapping approach \cite{Jain}.

The beauty of Haldane's treatment is that it provides an {\em unambiguous}
determination of what
$\beta$ statistics should be assigned to the excitations.
Noting that $N_L=3N_e-2+n$ states in the lowest Landau level give
$n$ qh for $\nu$ near but below 1/3, we can write
Eq.~(\ref{Nnqh}) as
\begin{equation}
N_n = {(N_L+4)/3 + (1-1/3)(n-1) \choose n}.
\label{Haldaneqh}
\end{equation}
{}From Eq.~(\ref{Haldaneqh})
we can read off, by comparison with Eq.~(\ref{Nn}), a Haldane
coefficient $\beta=1/3$ for qh near $\nu=1/3$.
(We have also verified that $\beta_{qh} = +1/5$ for qh near
$\nu=1/5$.)
Hence we confirm, using exact electronic eigenstates and standard
counting of the qh, the identification $\beta_{qh} = +1/m$
originally given by Haldane. This is, to our knowledge, the first
non-variational calculation of fractional statistics in the FQHE
\cite{bigwip}.

Haldane's scheme can be used to obtain further information from the
structure of the low-energy spectrum.
We use as a guide pseudo-wavefunctions of the form
Eq.~(\ref{wf}).  Such wavefunctions vanish at zero separation according
to the power law  $r^{\alpha}$ where $r=|z_i-z_j|$.  It is possible
to identify smaller subsets which vanish according to a higher power law,
{\it e.g.}, as $r^{2l+\alpha}$.  The number of such states with 2$l$
extra zeros is, by Eq.~(\ref{Nnl}), ${N_e+n-2l(n-1)\choose n}$.
We find that it is in fact possible to identify subsets of 2-qh states
with $2l$ extra zeros.
This enables us to estimate effective two-body interaction energies for
the excitations, which in turn
allows a clear evaluation of the stability
of certain hierarchy fractions.

We use a procedure for analyzing the multiplets which has been described
previously \cite{JohnsonMacDonald}. A slight generalization of this
procedure
enables us to identify
the multiplet structures of the subsets with $2l$ or more extra zeros.
Consider, for example, $n=2$
qh and $N_e=6$ electrons.  There are, by Eq.~(\ref{Nn}), 28 states lying
below the $V_1$ gap;  these break into multiplets with $L=0,2,4,6$.
Now, by Eq.~(\ref{Nnl}), there are 15 states
with 2 or more extra zeros in the effective qh wavefunction;
these turn out to have multiplets $L=0,2,4$.  In a similar
way we can identify the subsets with
$2l$ or more extra zeros; the results are given in Table~\ref{multiplet}
\cite{multnote}.

Quasiholes are expected to repel one another. Thus states with extra zeros,
which are further apart, are expected to be low in energy \cite{He}.
That is generally what we find (see Table~\ref{multiplet}).
The multiplets which by
our analysis should have extra zeros usually turn out to be either the
lowest-energy multiplets, or among the lowest.
These results give
semiquantitative information about the details of the qh
interactions which can explain features of the FQHE hierarchy.
To begin with, notice that a daughter of the $1/m$ level exists whenever
the ground state is nondegenerate.  This happens when the state with $2l$
extra zeros is a singlet ($N_n^{2l} = 1$), if that state is the ground state.
If this occurs for $2l=2$, then there is one state in which all of the qh's
avoid the minimum angular momentum ({\it i.e.}, in which they all have an
extra pair of zeros).  The resulting state is $\nu=2/7$.  The states
in which all qh's have $2l=4,6,\dots$ extra relative zeros are,
respectively, $\nu=4/13,6/19,\dots$.

This arithmetic suggests
which states might exist in the hierarchy picture
\cite{HalperinHierarchy,HaldaneHierarchy}.  Whether or not
they actually are stable is a question of energetics.
We assume the energy of qh states depends on
their interactions, which can
be described by effective pseudopotential parameters.  Suppose $\tilde{V}_{2l}$
is the pseudopotential parameter for a pair of qh with an extra two units
of RAM beyond the minimum required by statistics.
We estimate the $\tilde{V}_{2l}$
by considering the case of $n=2$ qh, since,
in this case, every eigenstate has the qh in a state of definite
RAM, and we can read off the $\tilde{V}_{2l}$ directly \cite{V2l}.
(These parameters have been estimated for qp, using a different method
which is based on trial wavefunctions, by B\'eran and Morf
\cite{BeranMorf}.)

Our results are shown in Fig.~\ref{Veff}.  For example,
for $N_e=6$ electrons and $n=2$ qh, the multiplets are in order
$\tilde{V}_6<\tilde{V}_2<\tilde{V}_4<\tilde{V}_0$;  the multiplet with six
extra zeros is the lowest in energy, that with no extra zeros the highest.
The most interesting feature is that the $\tilde{V}_{2l}$ are non-monotonic,
unlike the $V_l$ which describe the Coulomb interaction between
electrons.  In every case tested we see one feature in common:
$\tilde{V}_2<\tilde{V}_4$.  This non-monotonicity can explain the absence
of certain fractions.
The hierarchy
construction predicts that $\nu=4/13$ is a daughter of $1/3$
in which the qh avoid paying $\tilde{V}_2$, {\it i.e.}, every pair of qh
has at least four extra zeros.  By doing so, however, they pay a
$\tilde{V}_4$ cost.  If $\tilde{V}_4>\tilde{V}_2$, as we find, the
$\nu=4/13$ state is not stable---a conclusion apparently consistent
with experiment \cite{expts}.

We also note that, in contrast to the estimates of B\'eran and Morf
\cite{BeranMorf}
for qp, our estimates for $\tilde{V}_{2l}$ are not always positive.
These results are puzzling, but cannot be viewed as an artifact of our
method: they simply mean that there are some 2-qh states which are
lower in energy than twice the lowest 1-qh energy.  Further
study is needed to test the significance of this result.

The detailed analysis we have performed of the qh spectrum can be
repeated for the qp case, with the important difference \cite{ZX}
that the gap separating the low-lying states is typically of the same
magnitude as the splitting among these states---ie, of order $V_1$
[compare Figs.~\ref{spectrum}(a)) and
\ref{spectrum}(b)].
However, we can in most cases identify
(in agreement with Ref.~\onlinecite{He})
a low-energy subspace
with which to test Haldane counting.

Haldane's argument suggests that qp should have
statistics coefficient $\beta=-1/m$.
Note that qp see electrons as flux in the opposite direction of
that seen by qh, and so the appropriate pseudo-wavefunctions to
guide analysis of the qp spectrum involve factors of the
form \cite{HalperinHierarchy}
$(z_i^*  - z_j^*)^{-1/m}$. However, such a behavior for qp
at short distances is energetically costly \cite{He}.
Hence we expect not to see a subspace with negative $\beta$,
but rather only subsets with $\beta = 2l-(1/m)$ for nonzero
$2l$.

This expectation is borne out in our results.
Repeating the procedure applied to qh
states, we find
a Haldane coefficient $\beta=2-1/3$ for
qp states near $\nu=1/3$.  For general $\nu=1/m$, this approach gives
$\beta=2-1/m$.  As argued above, the states in the low-energy subspace
can be viewed as those of $\beta=-1/3$ particles
with 2 or more extra zeros.  In fact, either the
states with no extra zeros do not exist at all, or they are hidden
in the higher-energy states.  Hence it is better to
view $\beta=2-1/3$ as the fundamental statistics coefficient for the qp,
to be compared with $\beta=1/3$ for qh.

The multiplet structures can be also understood as in the qh case;
and the result of such analysis is strikingly similar.
We again find a consistently nonmonotonic behavior of the $\tilde{V}_{2l}$
($\tilde{V}_{4} > \tilde{V}_{2} $ always),
with $\tilde{V}_{2}$ and $\tilde{V}_{6}$ typically negative.
Hence we find, in close analogy to  our qh result, that
the 4/11 state should not be stable---again in agreement with
experiment \cite{expts} and with previous work \cite{GrosMacDonald,BeranMorf}.

In summary, we have tested the application
of Haldane's generalization of the Pauli
principle to charged excitations of the FQHE, using exact spectra for
small numbers of electrons, and assuming only that we know the
(fractional) charge for the
excitations.
At the largest energy scale at which the daughter states are defined,
we find fractional exclusion coefficients
$\beta_{qh} = 1/3$ and $\beta_{qp} = 2 - 1/3$.
This approach also permits us to find subsets described by coefficients
$\beta_{qh,qp}+2l$.
These states can be thought
of as pseudo-wavefunctions with extra (pairs of)
zeros in the relative coordinates.
Our analysis then enables us to estimate the effective
pseudopotential parameters for the excitations.
These turn out to be non-monotonic
with increasing relative angular momentum, which
then implies that certain daughter states ({\it e.g.},
$\nu = 4/13 $ and $\nu = 4/11$)
should not be stable incompressible states.

G.S.C. acknowledges partial support from the  NSF under
Grant \# DMR-9101542, and M.D.J. acknowledges partial support
by the UCF Division of Sponsored Research.  We acknowledge support from
Florida State University by the allocation of supercomputer resources.

\begin{table}\caption{ Multiplicity of eigenstates for many-quasihole
states near $\nu=1/3$.  $N_e$ is the number of electrons, $n$ the number
of quasiholes, and $N_L$ the corresponding number of states in the lowest
Landau level on the sphere. $N_n$ is the total number of quasihole states,
{\it i.e.}, those lying below the gap of order $V_1$.  The remaining columns
show the multiplicities of the degenerate states with no extra zeros,
two or more extra zeros, etc.  The multiplets are listed in order of
increasing energy.  Notice that states with extra zeros are typically
lower in energy.
\label{multiplet}}
\begin{tabular}{rrrrrrrr}
$N_e$ & $N_L$ & $n$ & $N_n$ & $2l\ge 0$ & $2l\ge 2$ & $2l\ge 4$ & $2l\ge 6$ \\
\tableline
4 & 12 & 2 & 15 & 5,9,1 & 5,1  & 1  &-  \\
  & 13 & 3 & 35 & 1,7,9,5,13 & 1  & - & - \\
  & 14 & 4 & 70 & 5,9,11,5,1,13,9,17 & - & - & - \\
5 & 15 & 2 & 21 & 7,3,11 & 7,3  & 3  & - \\
  & 16 & 3 & 56 & 4,10,8,12,6,16 & 4  & - & - \\
  & 17 & 4 & 126 & 5,9,7,13,$\dots$  & - & -  & - \\
  & 18 & 5 & 252 & 4,10,8,12,$\dots$ & - & - & - \\
6 & 18 & 2 & 28 & 1,9,5,13 & 1,9,5  & 1,5  & 1  \\
  & 19 & 3 & 84 & 7,13,3,11,15,7,9,19 & 7,3  & - & -  \\
  & 20 & 4 & 210 & 1,9,13,7,$\dots$ & 1  & -  & - \\
  & 21 & 5 & 462 & 7,15,13,9,$\dots$ & -  & - & - \\
  & 22 & 6 & 924 & 1,9,13,5,$\dots$ & -  & - & - \\
7 & 21 & 2 & 36 & 11,3,7,15 & 11,3,7  & 3,7  & 3  \\
  & 22 & 3 & 120 & 10,6,16,8,$\dots$ & 10,6,4  & - & - \\
\end{tabular}
\end{table}

\begin{figure}\caption{(a) The low-energy spectrum for 2 quasiholes
and 6 electrons on a sphere, in the truncated pseudopotential model
with $V_3=0.01V_1$.  The quasihole states lie below a gap of order $V_1$
and vary on an energy scale of order $V_3$.
(b) The low-energy spectrum for 3 quasiparticles and 8 electrons
on a sphere, under the conditions of (a).   The quasiparticle states lie
below a less evident gap of order $V_1$, and vary on an energy scale which
is also of order $V_1$.
\label{spectrum}}\end{figure}

\begin{figure}\caption{ Effective pseudopotential parameters for
quasiholes near $\nu=1/3$, as extracted from two-quasihole spectra.
$\tilde{V}_{2l}$ is the energy of a pair of quasiholes in a state
with relative angular momentum $2l$ more than the minimum required
by statistics.  Notice the non-monotonicity which can explain the absence
of certain FQHE fractions (see text).  The solid lines are for the
truncated pseudopotential model, with $V_3=0.01V_1$.  The dashed line
is the Coulomb result (in units of $e^2/\ell$, where $\ell$ is the
magnetic length).
We find similar results for quasiparticles.
\label{Veff}}\end{figure}


\begin{references}
\bibitem{bigwip} Given our identification of $\beta$ with $\alpha$
for the lowest Landau level (which is appropriate here), our
extraction of fractional $\beta$-statistics also implies fractional
$\alpha$-statistics.
\bibitem{multnote} We note that only for 2 qh does each multiplet
correspond to states with a definite number of extra zeros.
For example, with 3 qh and 6 electrons, there are 84 states below
the $V_1$ gap, in 8 different multiplets.  Of these, two multiplets
($L=1,3$) have two extra zeros.
\bibitem{V2l} We take $\tilde{V}_{2l}= (E_{2l}-E_0) - 2(E_1-E_0)$, where
$E_{2l}$ is the energy of the relevant 2-qh multiplet and $E_{0,1}$ is the
lowest-energy state with 0 or 1 qh present \cite{BeranMorf}.
\end{references}
\end{document}